\documentclass[slac_one]{revtex4}
\usepackage{graphicx}
\usepackage{fancyhdr}
\begin{document}

\title{Inclusive high \boldmath$Q^2$ cross sections and QCD and EW fits at HERA} 

\author{A. D. Tapper (on behalf of the H1 and ZEUS collaborations)}
\affiliation{Imperial College London, The Blackett Laboratory, Prince
Consort Road, London, SW7 2BW, U.K.}

\begin{abstract}
The latest measurements of the cross sections for neutral and charged current
deep inelastic scattering in $e^{\pm}p$ collisions at HERA with longitudinally
polarised lepton beams are presented. The measured cross sections are
compared with the predictions of the Standard Model. Next-to-leading-order 
QCD analyses of HERA data are also presented. 
The analyses yield the parton distribution functions
of the proton and are extended to also extract electroweak parameters. 
The determinations from HERA are compared with those from LEP and the Tevatron.
\end{abstract}

\maketitle

\thispagestyle{fancy}

\section{INTRODUCTION}
Deep inelastic scattering (DIS) of leptons off nucleons probes the
structure of matter at small distance scales. Two types of DIS
interactions are possible at HERA: neutral current (NC) reactions
$e^{-}p \rightarrow e^{-} X$ and $e^{+}p \rightarrow e^{+} X$, where a
photon or $Z^{0}$ boson is exchanged and charged current (CC)
interactions $e^{-}p \rightarrow \nu X$ and
$e^{+}p \rightarrow \bar{\nu} X$, where a $W^{\pm}$ boson is
exchanged.

The kinematics of charged current and neutral current deep inelastic
scattering processes are defined by the four-momenta of the incoming
lepton ($k$), the incoming proton ($P$), the outgoing lepton ($k'$)
and the hadronic final state ($P'$). The four-momentum transfer
between the electron and the proton is given by $q=k-k'=P'-P$. The
square of the centre-of-mass energy is given by $s=(k+P)^{2}$. The
description of DIS is usually given in terms of three Lorentz
invariant quantities, which may be defined in terms of the
four-momenta $k$, $P$ and $q$:
\begin{itemize}
\item $Q^{2}=-q^{2}$, the negative square of the four-momentum transfer,
\item $x=\frac{Q^{2}}{2P\cdot q}$, the Bjorken scaling variable,
\item $y=\frac{q\cdot P}{k\cdot P}$, the fraction of the energy transferred 
to the proton in its rest frame.
\end{itemize}
These variables are related by $Q^{2}=xys$, when the masses of the
incoming particles can be neglected.

This paper presents the most recent measurements of the cross sections for $e^\pm p$ NC
and CC DIS with longitudinally polarised lepton beams. The 
measurements are based on around $300~{\rm pb}^{-1}$ of data collected
at a centre-of-mass energy of $318~{\rm GeV}$. The measured
cross sections are compared to the Standard Model predictions.
Next-to-leading-order QCD analyses of HERA data are also
presented. The analyses are used to extract the parton distribution
functions (PDFs) of the proton and are extended to also extract electroweak
parameters.

\section{CROSS SECTIONS}
\label{s:xsects}
The longitudinal polarisation of the lepton beam is defined as
${P_{e}}=(N_{R}-N_{L})/(N_{R}+N_{L})$, where $N_{R}$($N_{L}$) are the numbers of right(left)-handed
leptons in the beam. The double-differential cross section for the neutral current process
with polarised lepton beams is given by 
\begin{equation}
\frac{d^2 \sigma ^{\rm NC} (e^\pm p)}{dxdQ^2}=\frac{2\pi \alpha ^2}{xQ^4} [H_{0}^{\pm}+P_{e}H_{P_{e}}^{\pm}], \nonumber
\end{equation}
where $\alpha$ is the QED coupling constant and $H_{0}^{\pm}$ and $H_{P_{e}}^{\pm}$ contain the unpolarised and polarised structure functions, respectively, such that at leading order in QCD
\begin{equation}
H_{0,P_{e}}^{\pm}=Y_{+}F_{2}^{0,P_{e}} \mp Y_{-}xF_{3}^{0,P_{e}}, \hspace{1cm} F_{2}^{0,P_{e}}=\sum_q x(q+\bar{q})A_{q}^{0,P_{e}}, \hspace{1cm} xF_{3}^{0,P_{e}}=\sum _q x(q-\bar{q})B_{q}^{0,P_{e}}, \nonumber
\end{equation}
where $Y_{\pm}=1\pm(1-y)^{2}$ and $q(x,Q^{2})$ and $\bar{q}(x,Q^{2})$
are the quark and antiquark PDFs, respectively, 
and the sums run over the five active
quark flavours. The $A$ and $B$ coefficients contain the quark and
lepton couplings to the photon and $Z^0$ boson and are given by
\begin{equation}
A_{q}^{0}=e_{q}^{2}-2e_{q}v_{q}v_{e}\chi_{Z}+(v_{q}^{2}+a_{q}^{2})(v_{e}^{2}+a_{e}^{2})\chi_{Z}^{2}, \hspace{1cm} B_{q}^{0}=-2e_{q}a_{q}a_{e}\chi_{Z}+4v_{q}a_{q}v_{e}a_{e}\chi_{Z}^{2}, \nonumber
\end{equation}
\begin{equation}
A_{q}^{P_{e}}=2e_{q}v_{q}a_{e}\chi_{Z}-2(v_{q}^{2}+a_{q}^{2})v_{e}a_{e}\chi_{Z}^{2}, \hspace{1cm} B_{q}^{P_{e}}=2e_{q}a_{q}v_{e}\chi_{Z}-2v_{q}a_{q}(v_{e}^{2}+a_{e}^{2})\chi_{Z}^{2}, \nonumber
\end{equation}
where $e_{f}$ is the electric charge in units of the positron charge
and $a_{f}$ and $v_{f}$ are 
the axial and vector couplings of the fermion $f$. The quantity
$\chi_{Z}$ is 
proportional to the ratio of the $Z^{0}$ and photon propagators
\begin{equation}
\chi_{Z}=\frac{1}{\sin ^{2} 2\theta_W} \bigg ( \frac{Q^{2}}{M_{Z}^{2}+Q^{2}} \bigg ), \nonumber
\end{equation}
where $M_{Z}$ is the mass of the $Z^{0}$ boson and $\theta_W$ is the Weinberg angle. 

The cross section for CC DIS with a longitudinally polarised lepton beam, can be expressed as
\begin{equation}
\frac{d^2 \sigma ^{\rm CC}(e^\pm p)}{dxdQ^2}=(1\pm P_{e})\frac{G_{F}^{2}}{4\pi x}\bigg(\frac{M_{W}^{2}}{M_{W}^{2}+Q^{2}}\bigg) ^{2}\bigg[ Y_{+}F_{2}^{\pm}-Y_{-}xF_{3}^{\pm} -y^{2}F_{L}^{\pm}\bigg], \nonumber
\end{equation}
where $G_{F}$ is the Fermi constant and $M_{W}$ is the mass of the $W^\pm$
boson. The structure functions $F_{2}^{\pm}$ and $xF_{3}^{\pm}$ contain sums and differences of the
quark and anti-quark PDFs and $F_{L}^{\pm}$ is the longitudinal structure function. 

\section{RESULTS}
\begin{figure*}[t]
\centering
\mbox{
  \includegraphics[width=.40\textwidth]{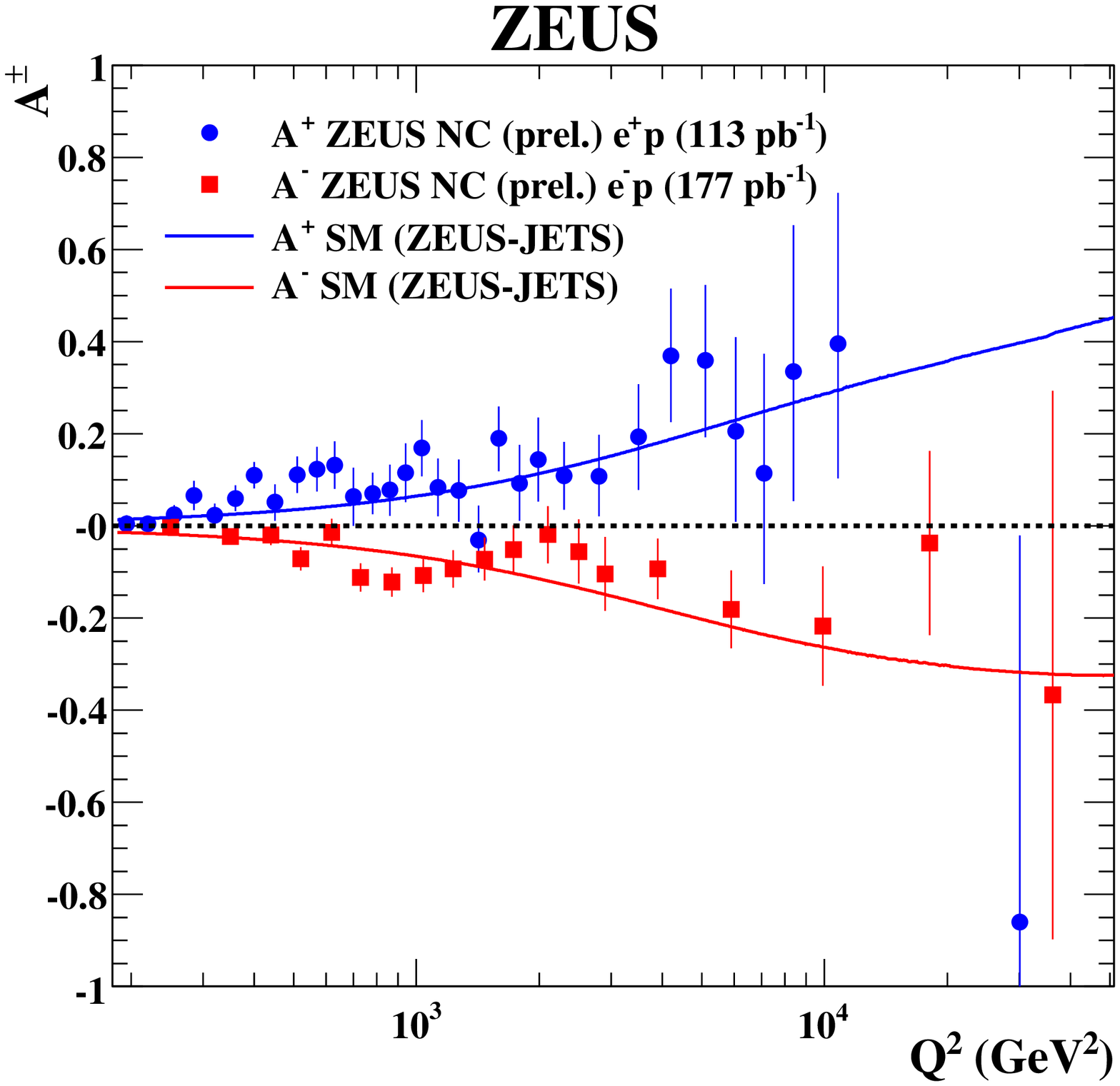}
  \includegraphics[width=.45\textwidth]{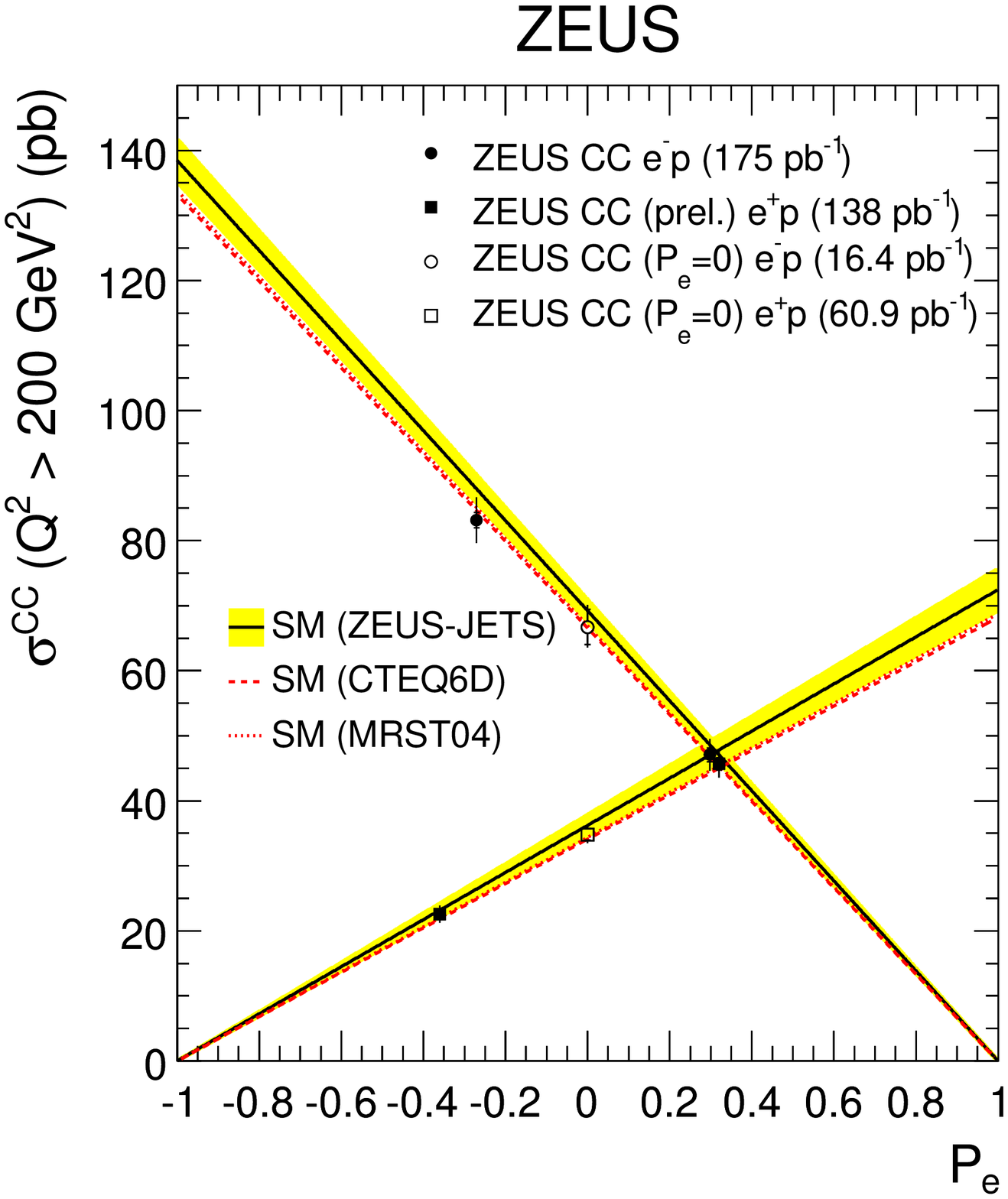}
}
\caption{The neutral current
  asymmetry $A^\pm$ as a function of $Q^{2}$ (left) and the cross
  section for charged current DIS as a function of
  the polarisation of the lepton beam (right).} 
\label{f:xsects}
\end{figure*}

In order to study the polarisation dependence of the NC DIS cross
sections the polarisation asymmetry may be formed by
\begin{equation}
A^\pm = \frac{2}{P_{e}^+ - P_{e}^-}\frac{\sigma^\pm(P_{e}^+) - \sigma^\pm(P_{e}^-)}{\sigma^\pm(P_{e}^+) + \sigma^\pm(P_{e}^-)}, \nonumber
\end{equation}
where $P_{e}^\pm$ are the mean polarisation values of the positively and negatively
polarised data samples and $\sigma^\pm$ are the measured $e^\pm p$ cross
sections. The asymmetry is shown in Fig.~\ref{f:xsects} (left) as a function of $Q^2$. It can be seen that the
values are non-zero and are well described by the
Standard Model prediction evaluated using the ZEUS-JETS~\cite{ZEUS-J}
PDFs, confirming the expected polarisation dependence. The asymmetry
is proportional to $a_e v_q$ and therefore a direct measurement of
parity violation at high $Q^2$.

The cross sections for $e^{\pm}p$ CC DIS are shown as a function of the longitudinal
polarisation of the lepton beam in Fig.~\ref{f:xsects} (right) including the
unpolarised ZEUS measurements from the 1998-2000 data~\cite{ZCC9899,ZCC9900}.
The data are compared to the Standard Model prediction evaluated using
the ZEUS-JETS, CTEQ6~\cite{CTEQ} and MRST04~\cite{MRST}
PDFs. The Standard Model predicts a linear dependence on $P_{e}$ with the cross section
becoming zero for right-handed (left-handed) electron (positron)
beams, due to the chiral nature of the Standard Model. The predictions describe the data well. 
\section{COMBINED ELECTROWEAK AND QCD FIT}
The neutral current cross sections give information on the quark couplings to the
$Z^0$ boson which appear in the coefficients $A$ and $B$ in
Sect.~\ref{s:xsects}. One can see that since $v_e$ is small and 
$\chi_Z \gg \chi_Z^2$ for the HERA kinematic regime, the axial and
vector couplings are dominant in the unpolarised $xF_3$ and
polarised $F_2$ terms, respectively. These electroweak parameters can be fitted simultaneously 
with the PDF parameters to perform a model-independent
extraction~\cite{H1EW}. The details of the fits for the PDFs follow
 the procedures in previous H1 and ZEUS
publications~\cite{H1QCD,ZEUS-J} closely. The H1 fit includes inclusive NC and
CC DIS data over a large range in $Q^2$ and $x$. The ZEUS fit also
includes jet cross-section data. These fits are extended to extract
the quark couplings to the $Z^0$ boson from the high-$Q^2$ NC DIS data.
 
\begin{figure*}[t]
\includegraphics[width=.45\textwidth]{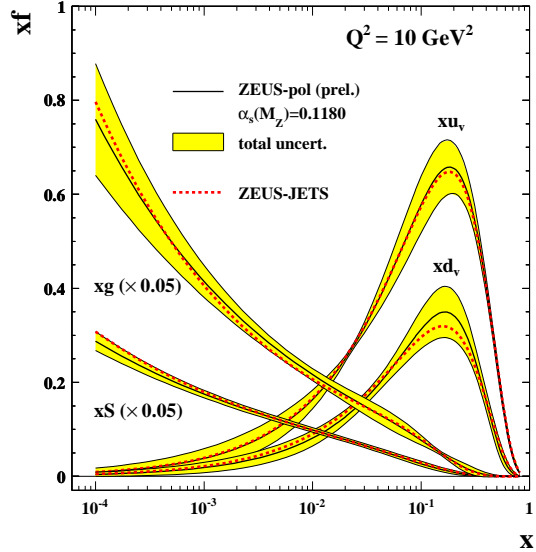}
\caption{The extracted parton density functions of the proton at $Q^2= 10~\rm{GeV}^2$. } 
\label{f:pdfs}
\end{figure*}

\begin{figure*}[t]
\centering
\mbox{
  \includegraphics[width=.45\textwidth]{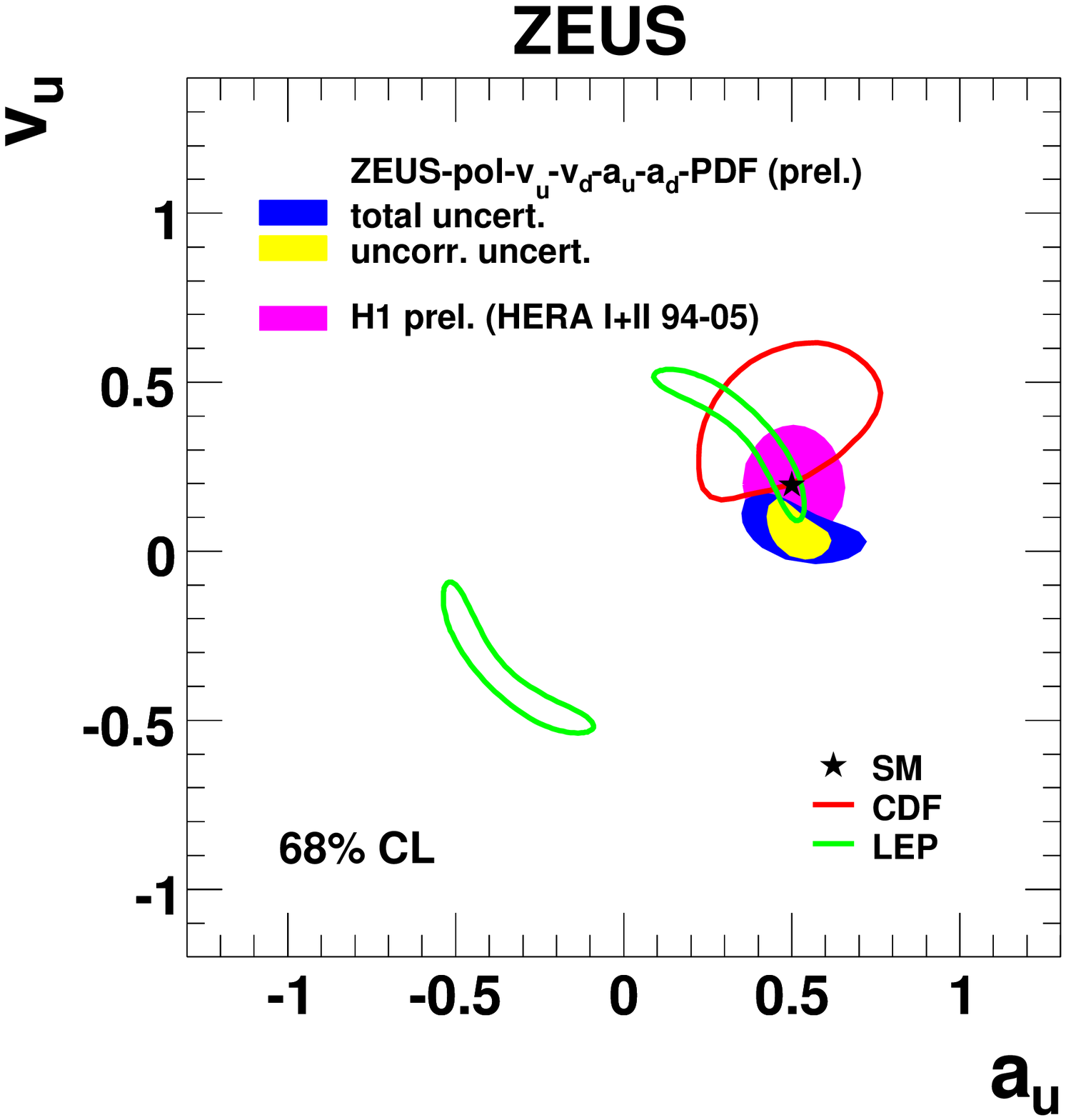}
  \includegraphics[width=.45\textwidth]{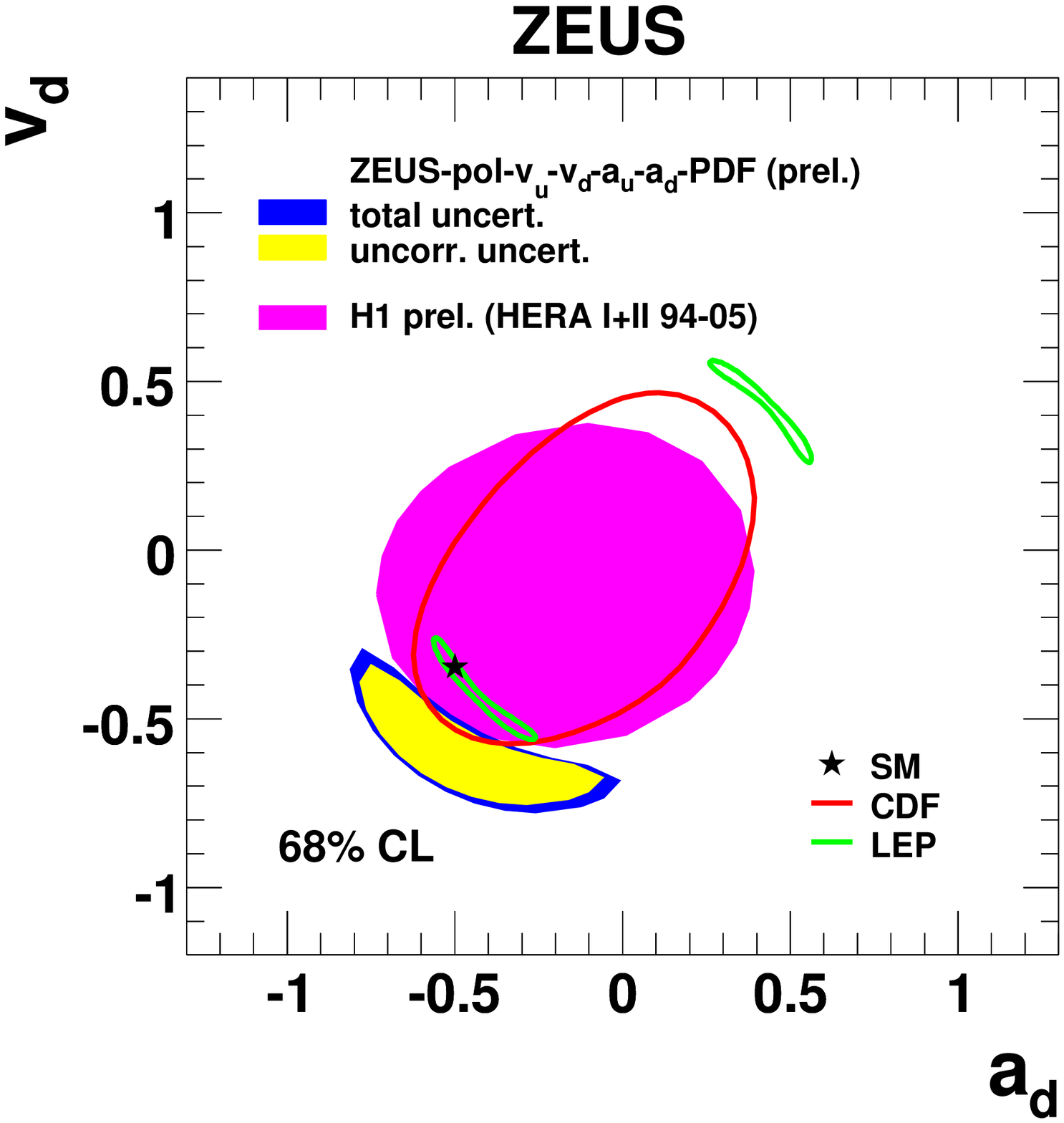}
}
\caption{The extracted values of the axial and vector couplings of the
$Z^0$ boson to the up (left) and down (right) quarks. The contours
represent 68\% confidence limits.} 
\label{f:couplings}
\end{figure*}

Figure~\ref{f:pdfs} shows the extracted PDF parameters compared to
those from the ZEUS-JETS fit.The PDFs are not significantly altered by
the inclusion of electroweak parameters in the fit. 
Figure~\ref{f:couplings} shows the results of the determination of $a_q$
and $v_q$ for $u$ and $d$ type quarks, together with results from
other experiments. The HERA data are more sensitive to light quarks
and thus the extracted couplings are compared with the corresponding 
light-quark extractions from LEP~\cite{LEP} and CDF~\cite{CDF}. 
The extracted values of the couplings are in agreement with
the Standard Model predictions and the precision is competitive with
measurements from other colliders.
\section{SUMMARY}
The cross sections for neutral and charged current deep inelastic
scattering in $e^\pm p$ collisions with longitudinally polarised
lepton beams have been measured. The latest measurements, based on around
$300~\rm{pb^{-1}}$ of data are well described by the Standard
Model. Combined QCD and electroweak fits to HERA data yield
determinations the parton distribution functions and simultaneously allow the
extraction of the light-quark couplings to the $Z^0$ boson from the 
high-$Q^2$ NC DIS data. The extracted values of the couplings are in agreement with
the Standard Model predictions and the precision is competitive with
measurements from other colliders.


\begin{thebibliography}{9}   

\bibitem{ZEUS-J}
ZEUS Coll., S. Chekanov et al., Eur. Phys. J C42, 1 (2005).

\bibitem{ZCC9899}
ZEUS Coll., S. Chekanov et al., Phys. Lett. B 539, 197 (2002). Erratum in Phys. Lett. B 552, 308 (2003).

\bibitem{ZCC9900}
ZEUS Coll., S. Chekanov et al., Eur. Phys. J. C 32, 1 (2003).

\bibitem{CTEQ}
J. Pumplin et al., JHEP 2027, 012 (2002).

\bibitem{MRST}
A.D. Martin et al., Eur. Phys. J C23, 73 (2002).

\bibitem{H1EW}
H1 Coll., A. Aktas et al., Phys. Lett. B 632, 35 (2006).

\bibitem{H1QCD}
H1 Coll., C. Adloff et al., Eur. Phys. J. C 30, 1 (2003).

\bibitem{LEP}
LEP Electroweak Working Group, Phys. Rep. 427, 257 (2006).

\bibitem{CDF}
CDF Coll., D. Acosta et al., Phys. Rev. D71, 052002 (2005).

\end{thebibliography}
\end{document}